\documentclass[twocolumn,amsmath,amssymb,prl,superscriptaddress]{revtex4-1}
\pdfoutput=1

\usepackage{graphicx}% Include figure files
\usepackage{dcolumn}% Align table columns on decimal point
\usepackage{bm,psfrag}% bold math
\usepackage[utf8]{inputenc}

\newcommand{\be}{\begin{equation}}
\newcommand{\ee}{\end{equation}}
\newcommand{\bea}{\begin{eqnarray}}
\newcommand{\eea}{\end{eqnarray}}
\newcommand{\ba}{\begin{eqnarray}}
\newcommand{\ea}{\end{eqnarray}}

\newcommand{\beq}{\begin{equation}}
\newcommand{\eeq}{\end{equation}}
\newcommand{\beqa}{\begin{eqnarray}}
\newcommand{\eeqa}{\end{eqnarray}}
\newcommand{\beqar}{\begin{eqnarray*}}
\newcommand{\eeqar}{\end{eqnarray*}}
\newcommand{\e}{\epsilon}

\newcommand{\ie}{{\it i.e.,}\ }

\newcommand{\eqlabel}[1]{\label{#1}}  %{\quad\mt{#1}\label{#1}}   %

\def\tr{\rm tr}

\def\t6 {T_\mt{D6}}

\newcommand{\mt}[1]{\textrm{\tiny #1}}

\def\cale         {{\cal E}}

\def\call         {{\cal L}}

\def\caln         {{\cal N}}
\def\calo         {{\cal O}}

\def\calt         {{\cal T}}

\def\del          {\partial}

\def\ee           {{\rm e}}

\def\tr           {\mathop{\rm Tr}}

\def\Im           {{\rm Im\hskip0.1em}}

\def\sqr#1#2{{\vcenter{\vbox{\hrule height.#2pt
 \hbox{\vrule width.#2pt height#1pt \kern#1pt
 \vrule width.#2pt}\hrule height.#2pt}}}}

\def\a{\alpha}

\def\w{\omega}

\def\dd{\delta}
\def\e{\epsilon}
\def\c{\chi}

\def\ee{\cale}

\def\aa1{\phi}
\def\cc1{\psi}

\newcommand{\eq}{\begin{equation}}
\newcommand{\eqx}{\end{equation}}

\usepackage[bookmarks=false]{hyperref} %make sure it comes last of your loaded packages
\hypersetup{pdfstartview=FitH,pdfhighlight=/O,colorlinks=false}

\begin{document}

%\preprint{arXiv:1603.nnnn [hep-th]}

%\title{Beyond adiabatic approximation in Big Bang Cosmology:\\ hydrodynamics, resurgence and entropy production in the Universe}
%has zero radius of convergence

% J: Perhaps we should change the title? The issue is that the current title is much more general than we actually do and may have given the wrong idea to the referee (who probably thought about earlier times in the Universe when T was very large)

\title{Entropy Production, Hydrodynamics, and Resurgence \\ in the Primordial Quark-Gluon Plasma from Holography}

\author{Alex Buchel}
\email{abuchel@perimeterinstitute.ca}
\affiliation{Department of Applied Mathematics, Department of Physics and Astronomy, University of Western
Ontario, London, Ontario N6A 5B7, Canada}
\affiliation{Perimeter Institute for Theoretical Physics, Waterloo, Ontario N2L 2Y5,
Canada}
\author{Michal P. Heller}
\email{mheller@perimeterinstitute.ca}
\altaffiliation[On leave from: ]{\emph{National Centre for Nuclear Research,  Ho{\.z}a 69, 00-681 Warsaw, Poland.}}
\affiliation{Perimeter Institute for Theoretical Physics, Waterloo, Ontario N2L 2Y5,
Canada}
\author{Jorge Noronha}
\email{noronha@if.usp.br}
\affiliation{Instituto de F\'{\i}sica, Universidade de S\~{a}o Paulo, C.P. 66318,
05315-970 S\~{a}o Paulo, SP, Brazil}

\begin{abstract}
Microseconds after the Big Bang quarks and gluons formed a strongly-coupled non-conformal liquid driven out-of-equilibrium by the expansion of the Universe. We use holography to determine the non-equilibrium behavior of this liquid in a Friedmann-Lemaitre-Robertson-Walker Universe and develop an expansion for the corresponding entropy production in terms of the derivatives of the cosmological scale factor. We show that the resulting series has zero radius of convergence and we discuss its resurgent properties. Finally, we compute the resummed entropy production rate in de Sitter Universe at late times and show that the leading order %adiabatic 
approximation given by bulk viscosity effects can strongly overestimate/underestimate the rate depending on the microscopic parameters.

 %for massive gauge theory plasma
%nonconformal matter in FLRW Universe has zero radius of convergence. We compute the resummed entropy production rate in de Sitter Universe 
%at late times and show that the leading order adiabatic approximation can strongly overestimate/underestimate the rate, depending on the microscopic parameters.
%%for plasma with fermion/boson masses the
%hydrodynamic approximation strongly overestimate/underestimate the rate. We generalize the results for a strongly 
%coupled conformal gauge theory plasma, perturbed by a relevant operator of dimension $\Delta\ge 2$.  
\end{abstract}

\maketitle

\noindent {\it Introduction.---} While the Quark-Gluon Plasma (QGP) formed at the and LHC has been under intense investigation for more than a decade \cite{Heinz:2013th}, much less is known about the dynamical (i.e., real time) non-equilibrium processes that took place in the early Universe when its temperature was close to the QCD crossover transition ($T \sim 150-300$ MeV) \cite{Aoki:2006we}. Under these conditions, QCD matter behaved as a strongly interacting, non-conformal nearly perfect fluid driven out-of-equilibrium by the rapid expansion of the Universe. 

In this article we propose to use the holographic correspondence \cite{Maldacena:1997re,Witten:1998qj,Gubser:1998bc} to study strongly coupled non-equilibrium phenomena in an expanding Universe at temperatures near the QCD phase transition, which are currently beyond the scope of other non-perturbative approaches such as lattice gauge theory (for $T \gg 300$ MeV the QGP becomes weakly coupled and its dynamics should be well described by kinetic theory \cite{Arnold:2002zm}). In this strong coupling regime, the holographic correspondence is the only available tool to study real time dynamics in non-Abelian gauge theories. We use the deconfined phase of strongly interacting $\caln=2^*$ gauge theory, a close cousin of the strongly coupled QGP with a well-known holographic description, and solve for its dynamics in Friedmann-Lemaitre-Robertson-Walker (FLRW) backgrounds 
using holography. We investigate the entropy produced in this system by the expansion of the Universe and show that for this cosmological fluid the relativistic hydrodynamic expansion, the standard tool to investigate near equilibrium behavior, is an asymptotic series with zero radius of convergence. The study presented in this article is part of the recent ongoing efforts towards understanding the emergence of relativistic hydrodynamics both at strong \cite{Heller:2011ju,Chesler:2015fpa,Rangamani:2015qga} and weak coupling \cite{Denicol:2014xca,Denicol:2014tha,Kurkela:2015qoa,Bazow:2015dha} and its behavior in the presence of large gradients.

The standard model of cosmology \cite{Weinberg:2008zzc} is based upon the (spatially) maximally symmetric FLRW line element with zero spatial curvature
\eq
\label{b.metric}
ds^2 =g_{\mu\nu}dx^\mu dx^\nu=-dt^2+a^2(t) \, d\boldsymbol{x}^2%\left(dx^2+dy^2+dx^2\right)
,\eqx
where the cosmological scale factor $a(t)$ is governed by Einstein's equations coupled to matter
\eq
\label{einst}
R_{\mu\nu}-\frac 12 R g_{\mu\nu}=8\pi G\, T_{\mu\nu}.
\eqx
%hypothesis of homogeneity and isotropy
Under these symmetry assumptions, the matter stress tensor
is diagonal in the coordinates \eqref{b.metric} and it reads $T_{\mu\nu}={\rm diag}\{\e,P,P,P\}$,
%\eq\label{tmatter}
%T_{\mu\nu}={\rm diag}\{\e,P,P,P\},
%\eqx 
where $\e$ is the energy density of the system and $P$ is the pressure. To close the equations of motion for the scale factor, one necessarily needs an equation 
relating the energy density and the pressure. Typically, one assumes an equilibrium 
relation: $P=P(\e)$. An immediate consequence of this choice is that the 
expansion is an adiabatic (reversible) process --- in this case no entropy was produced by the strongly coupled QGP liquid that once filled the expanding Universe.

Here we explore what it takes to go beyond this idealization. In the FLRW Universe the matter expansion is locally static 
$u^{\mu}=(1,0,0,0)$ though it possesses a nonzero expansion rate $\Theta\equiv \nabla_\mu u^\mu
=3\dot a/a$. In the hydrodynamic approximation the matter stress tensor can be written as a gradient expansion \cite{Romatschke:2009kr}
\eq\label{decompose}
T_{\mu\nu}=T_{\mu\nu}^{eq}+\Pi_{\mu\nu}(\dot a,\{\dot{a}^2,\ddot a\},\cdots ),
\eqx
where $\Pi_{\mu\nu}$ represents the derivative corrections to the equilibrium energy-momentum tensor $T_{\mu\nu}^{eq}$. In the Landau-Lifshitz frame \cite{Landau:1980mil} the lowest order correction is
\eq\label{hydro}   
\e=\e^{eq}+\calo\left(\dot{a}^2,\ddot a\right)\,,\ P=P^{eq}-\zeta (\nabla\cdot u)+\calo\left(\dot{a}^2,\ddot a\right) ,
\eqx
where $\zeta$ is the bulk viscosity \footnote{There is no contribution in \eqref{hydro} from the shear viscosity since the 
shear tensor vanishes identically in FLRW.}. The leading order expression for the entropy production rate can be obtained \cite{Landau:1980mil} from the stress-energy conservation $u^\nu \nabla^\mu T_{\mu\nu}=0$ combined with the
definition of the entropy density near-equilibrium $s|_{hydro} \approx (\e+P)/T$:
\eq\label{enthydro}
\frac{d}{dt}\ln \left(a^3 s\right)\bigg|_{hydro} \approx \frac 1T \left(\nabla\cdot u\right)^2\ \frac{\zeta}{s}.  
\eqx
For matter invariant under conformal transformations $\zeta=0$ and the above expression is zero. In fact, in this case there is no entropy production at any order in the hydrodynamic derivative expansion~\eqref{decompose}. One thus faces several challenges in understanding the entropy production of strongly interacting matter near the QCD crossover transition in an FLRW Universe. First and foremost, any model employed must not be conformal in order to correctly assess the degree of conformal symmetry breaking observed in QCD \cite{Aoki:2006we} and have long-lasting dissipative effects; second, it should be possible to reliably compute $T_{\mu\nu}$ at strong coupling beyond the leading order hydrodynamic approximation characterized by \eqref{hydro}; and last, but not least, the framework must possess a natural definition for the entropy density away from equilibrium. The holographic $\caln=2^*$ plasma fulfills the criteria listed above.  

\vspace{10 pt}

\noindent {\it $\caln=2^*$ gauge theory in FLRW.---} $\caln=2^*$ gauge theory arises as a relevant deformation of conformal $\caln=4$ Super Yang-Mills (SYM) theory by adding masses to some of its bosonic and fermionic fields \footnote{We refer the interested reader to the Appendix where we present a more detailed discussion about $\caln=2^{*}$ gauge theory.}. This corresponds, respectively, to a relevant deformation with $\Delta=2$ and $\Delta=3$ scalar operators schematically denoted by ${\cal O}_{b}$ and ${\cal O}_{f}$. The holographic description of $\caln=2^*$ gauge theory at strong coupling is given by the effective 
five-dimensional action \cite{Pilch:2000ue}
\eq
\label{eq.Lgrav}
S = \frac{1}{16 \pi G_{5}} \int \mathrm{d}^{5} x \sqrt{- g} \left(R 
-12 (\partial \alpha)^{2} - 4 (\partial \chi)^{2} - V 
 \right),
\eqx
with the scalar potential taking the form
\eq
\label{eq.VPW}
V =   -  e^{-4 \alpha} - 2 e^{2 \alpha} \cosh{2 \chi} + \frac 14 e^{8 \alpha} \sinh^{2}{2 \chi}.
\eqx
The gravitational bulk scalars 
$\a$ and $\chi$ represent, holographically, the dimension
$\Delta=2$ and $\Delta=3$ operators (the bosonic $m_b$ 
and the fermionic $m_f$ mass terms). The 
five-dimensional Newton's constant is $G_5=4\pi/N^2$ (we set the asymptotic 
AdS$_5$ radius $L=2$). Even though the matter content of this theory differs from QCD, we note that for $T \gtrsim 300$ MeV the breaking of conformal invariance in $\caln=2^*$ gauge theory is a reasonable approximation to the corresponding result in QCD \cite{Buchel:2015ofa}. 

We use the characteristic formulation of gravitational dynamics in asymptotically AdS spacetimes summarized in \cite{Chesler:2013lia} to describe 
non-equilibrium states of $\caln=2^*$ gauge theory in FLRW. Assuming homogeneity and isotropy in the spatial boundary coordinates $\boldsymbol{x}=\{x,y,z\}$ leads to the 5-dimensional metric Ansatz   
\begin{equation}
ds_5^2=2 dt\ (dr -A dt) +\Sigma^2\ d\boldsymbol{x}^2,
\eqlabel{EFmetric}
\end{equation}
where the warp factors $A,\Sigma$ as well as the bulk scalars $\a$ and $\chi$ depend only on $\{t,r\}$. 
The near-boundary $r\to\infty$ 
asymptotic behavior of the metric and the scalars encode the 
mass parameters $m_b$ and $m_f$ of the $\caln=2^*$ gauge theory \cite{Buchel:2007vy} 
and the boundary metric \eqref{b.metric} scale factor $a(t)$: 
\begin{eqnarray}
&&\Sigma=\frac{a}{r}+\calo(r^{-1})\,,\ A=\frac{r^2}{8}-\frac{\dot a r}{a }+\calo(r^0)\nonumber\\
&&\a=-\frac{8m_b^2\ln r}{3 r^2}+\calo(r^{-2})\,,\ \chi=\frac{2m_f}{r}+\calo(r^{-2}).
\label{bcdata}
\end{eqnarray}
The subleading terms in the near-boundary expansion of the 
metric functions and the scalars encode the non-equilibrium energy density $\e$,
pressure $P$, and the expectation values of the operators $\calo_b$ 
and $\calo_f$. Explicit formulas for these observables can be found in the Appendix.

An advantage of the holographic formulation is the natural definition of the 
dynamical, far-from-equilibrium entropy density at strong coupling: following \cite{Booth:2005qc,Figueras:2009iu} we associate the non-equilibrium entropy density 
%of the expanding $\caln=2^*$ plasma 
$s$ with the Bekenstein-Hawking entropy of the apparent horizon in the geometry \eqref{EFmetric},  
$a^3 s= N^2\Sigma^3/(16\pi) |_{r=r_h} $,
where $r_h$ is the location of the apparent horizon determined  from 
$d_+\Sigma|_{r=r_h}=0$ with $d_{+} \equiv \partial_{t} + A \, \partial_{r}$, see Ref.~\cite{Chesler:2013lia}.
Taking the derivative of the entropy density and using holographic equations of motion we find
\begin{eqnarray}
\label{erate}
&&\frac{d(a^3 s)}{dt}=\frac{2N^2}{\pi}\ (\Sigma^3)' \  
\frac{ (d_+\chi)^2+3 (d_+\a)^2}{-4 V}\bigg|_{r=r_h}.
%\notag
\end{eqnarray}
Numerical simulations are required to determine the properties of the 
$\caln=2^*$ gauge theory in FLRW for generic values of the masses \eqref{bcdata}; 
in what follows we consider the limit when the mass parameters $m_b$ and 
$m_f$ are much smaller than the local temperature $T$. This approximation captures all the essential physics and at the same time allows for the semi-analytic calculations. When $m_b=m_f=0$, the case of the parent conformal SYM in FLRW \cite{Apostolopoulos:2008ru}, $d(a^3 s)/dt=0$ and entropy production is zero to all orders in the derivative expansion. 

To proceed beyond scale invariance, we linearize the
bulk scalar equations $\phi_\Delta=\{\a,\chi\}$, $\Delta=\{2,3\}$
correspondingly, 
on the AdS-FLRW-Schwarzschild solution (see Eq.~\eqref{sym} in the Appendix): 
\begin{eqnarray}
&&d_+' \phi_\Delta+\frac{3}{2r}d_+\phi_{\Delta}+\frac{3}{16}\left(r-\frac{\mu^4}{a^4r^3}\right)\phi_\Delta' +
\nonumber \\ &&+ \frac{(4-\Delta)\Delta}{8}\phi_\Delta=0,
\label{phidelta}
\end{eqnarray}
where $d_+$ is computed with $A$ from Eq.~\eqref{sym} and $\mu$ is given by Eq.~\eqref{eq.tempexpand}. We solve Eq.~\eqref{phidelta} with $\phi_\Delta(t,r)$ satisfying the boundary conditions defined by Eq.~\eqref{bcdata} and being regular for $r\in (r_h\equiv\mu/a,\, +\infty)$. Although we are not directly interested in the dynamics of expectation values of ${\cal O}_{b}$ and ${\cal O}_{f}$ operators, solving Eq.~\eqref{phidelta} is the crucial point of this article as, via Eq.~(\ref{erate}), it provides us the entropy production rate.

%Although we are not directly interested in the dynamics of expectation values of ${\cal O}_{b}$ and ${\cal O}_{f}$ operators, s

\vspace{10 pt}

\noindent {\it Leading order hydrodynamic approximation.---} In the hydrodynamic regime, the Knudsen number $K_N =\Theta/T \sim \dot a/\mu\ll 1$ and the solution of Eq.~\eqref{phidelta} can be written as 
\eq\label{genexp}
\phi_\Delta=\dd_\Delta\ a^{4-\Delta} (4-\Delta)\sum_{n=0}^\infty \frac{\calt_{n,\Delta}[a]}{\mu^n}\ 
F_{\Delta,n}\left(\frac{\mu}{ar}\right),
\eqx
where $\calt_{\Delta,0}=\frac{1}{4-\Delta}$, and for $n>0$, $\calt_{\Delta,n}[a]$ is given by 
\eq\label{defrec}
\calt_{\Delta,n}=a\ \dot\calt_{\Delta,n-1}+(4-\Delta)\ \dot a\ \calt_{\Delta,n-1} .
\eqx
The radial profiles $F_{\Delta,n}(z)$ satisfy the nested system:
\begin{eqnarray}
&&0=F_{\Delta,n}''+\frac{(z^4+3)F_{\Delta,n}'}{z(z^4-1)}+\frac{(\Delta-4)\Delta F_{\Delta,n}}{z^2(z^4-1)}\nonumber\\
&&+\frac{8}{z^4-1}\left(
F_{\Delta,n-1}'-\frac{3}{2z} F_{\Delta,n-1}\right)\,,\qquad  n>0
\label{radial}
\end{eqnarray}
and $F_{\Delta,0}(z)=z^\Delta\ _2F_1(\Delta/4,\Delta/4;1;1-z^4)$.
For a $\caln=2^*$ gauge theory plasma \cite{Buchel:2007vy}
\eq\label{deltabf}
\dd_2=-\frac{2\pi m_b^2}{3\mu^2}\,,\qquad \dd_3=\frac{2\left[\Gamma\left(\frac34\right)\right]^2 m_f}{\pi^{1/2} \mu}.
\eqx
From Eq.~\eqref{erate}, the leading order (in the small mass expansion) contribution of the operator of dimension $\Delta$ to the entropy production 
reads 
\eq
\label{deltarate}
\frac{d(a^3s)}{dt}=\frac{N^2}{16\pi}a^{7-2\Delta}\mu^2\dd_\Delta^2(4-\Delta)^2s_\Delta \times \Omega_{\Delta}^{2},
\notag
\eqx
where $s_\Delta$ accounts for the normalization of the $\phi_\Delta$ kinetic term in the holographic action,
$s_3=1$ and $s_2=3$ in \eqref{eq.Lgrav}, and we defined
\eq
\Omega_{\Delta} \equiv \sum_{n=0}^\infty \calt_{\Delta,n+1}[a]\  \frac{F_{\Delta,n}(1)}{\mu^n}.
\eqx
Comparing Eq.~\eqref{deltarate} in the hydrodynamic approximation with Eq.~\eqref{enthydro}, i.e. keeping in Eq.~\eqref{deltarate} only the 
$F_{\Delta,n}(1)$ term, we recover bulk viscosity of $\caln=2^*$ plasma in the small-mass expansion~\cite{Buchel:2008uu} 
\eq\label{n2bulk}
\frac{\zeta}{s} =\frac{\left[\Gamma\left(\frac 34\right)\right]^4}{18\pi^4}\ \left(\frac{m_f}{T}\right)^2
+\frac{1}{216\pi^3}\ \left(\frac{m_b}{T}\right)^4.%+\calo\left(\frac{m_f^4}{T^4}\,,\frac{m_b^8}{T^8} \right)
\eqx
This serves as a powerful cross-check of our approach.

%Results of the computations of $F_{\Delta,n}$ for $\Delta=\{2,3\}$ for $n=1\cdots 100$ are presented in Fig.~\ref{figure1}. 

\vspace{10 pt}

\noindent {\it Hydrodynamics at large orders and resurgence.---} Quite remarkably, the recursion relation \eqref{defrec} can be solved in a closed form. For single-component cosmologies, $a(t)=\left(\frac{t}{t_0}\right)^{\frac{2}{3+3\w}}$, with constant $\w$ and $H\equiv \dot a/a$, we obtain
\eq
\calt_{\Delta,n}[a]=
\left(-\frac 12-\frac{3\w}{2}\right)^n\ \Gamma\left(n+\frac{2(\Delta-4)}{1+3\w}\right)\ a^nH^n.
\label{tncompute}
\eqx
Due to the factorial growth of the coefficient $\calt_{\Delta,n}$ at large~$n$, the hydrodynamic expansion has zero radius of convergence provided $F_{\Delta,n}(1)$ does not vanish too fast with~$n$. To see that this is not the case, we solved numerically Eqs.~\eqref{radial} for $\Delta = 3$ up to $n = 300$.

The ${\cal N}=2^{*}$ plasma in FLRW geometry \eqref{b.metric} is a \emph{genuinely new} kind of relativistic flow in which the hydrodynamic gradient expansion diverges. The only other known example \cite{Heller:2013fn,Heller:2015dha,Basar:2015ava,Aniceto:2015mto,Florkowski:2016zsi,Denicol:2016bjh,Heller:2016rtz} is the so-called Bjorken expansion relevant for the QGP dynamics in heavy-ion collisions~\cite{Bjorken:1982qr}. In the present situation it is the expanding spatial volume~\eqref{b.metric} that drives the system's dynamics due to its broken conformal symmetry, as opposed to the shear strain responsible for the dissipative nature of the Bjorken flow explored in \cite{Heller:2013fn,Heller:2015dha,Basar:2015ava,Aniceto:2015mto}. Thus, our results probe a completely different set of contributions to the stress energy tensor in the hydrodynamic regime.

Recently, hydrodynamics has been recognized as a fruitful new ground \cite{Heller:2015dha,Basar:2015ava,Aniceto:2015mto} for exploring ideas related to resurgence, i.e. the information carried by asymptotic expansions about their non-perturbative completion and the means of their resummation (see also \cite{Lublinsky:2007mm} for studies about the role of resummed hydrodynamic gradient expansions in heavy ion collisions). This is a very active area of research which sheds new light on the structure of quantum mechanical systems, see e.g.~\cite{Aniceto:2011nu,Argyres:2012vv,Argyres:2012ka,Dunne:2012ae,Aniceto:2013fka,Cherman:2014ofa}. The ability to solve the recursion relation~\eqref{tncompute} exactly, as well as Eq.~\eqref{radial} numerically with (in principle) arbitrarily high precision, allows us to bring new insights on the resurgent structure of hydrodynamic theories.

For simplicity, we focus on de Sitter cosmology
\eq
a = e^{H \, t},
\eqx
which can be obtained from Eq.~\eqref{tncompute} by taking the limit \mbox{$\w\to -1$}. In this case we obtain
\eq
\calt_{\Delta,n}[a]=\Gamma(n+4-\Delta) a^nH^n.
\eqx
Let us now investigate in detail the properties of $\Omega_{\Delta}$ defined in Eq.~\eqref{deltarate}, i.e. (the square root of) the contribution to the entropy production from the operator of dimension $\Delta$. It schematically reads
\eq
\Omega_{\Delta} = \sum_{n = 0}^{\infty} c_{n} g^{n},
\label{eq.seriesres}
\eqx
where we defined
\eq
\hspace{-4 pt}
c_{n} \equiv \frac{ \Gamma(n+4-\Delta) F_{\Delta,n}(1)}{\left( 4 \pi \right)^{n}} \,\,\, \mathrm{and} \,\,\, g \equiv \frac{H}{T} = \frac{4\pi}{\mu} a H.
\eqx
The hydrodynamic series may be viewed as a perturbative expansion in $g \ll 1$ valid for moderate times. The standard procedure in dealing with a divergent series such as~\eqref{eq.seriesres} is the technique of Borel transform  %it would be good to cite one of the resurgence papers on this spot%
\eq
\Omega_{\Delta}^{(B)} (\xi) = \sum_{n = 0}^{\infty} \, \frac{c_{n}}{n!} \xi^{n}
\eqx
and Borel resummation
\eq
\hspace{-5 pt} \Omega_{\Delta}^{(R)} = \int_{{\cal C}} d\xi \, e^{-\xi} \, \Omega^{(B)}_{\Delta} (\xi \, g) \equiv \frac{1}{g} \int_{{\cal C}} d\xi \, e^{-\xi/g} \, \Omega^{(B)}_{\Delta} (\xi),
\label{eq.OmegaR}
\eqx
where the contour ${\cal C}$ connects $0$ and $\infty$. The analytic continuation of the Borel transform, performed using Pad{\'e} approximants for truncated series such as \eqref{eq.seriesres}, reveals at least the leading (i.e. closest to the origin on the complex Borel plane) singularity responsible for the zero radius of convergence of the original series. The presence of such singularities make the Borel summation ambiguous (via the choice of contour ${\cal C}$), which requires introducing additional terms to the series to cancel the ambiguities. In hydrodynamic theories undergoing Bjorken expansion, the position of the singularity closest to the origin -- the beginning of a branch cut -- has been associated in \cite{Heller:2013fn,Heller:2015dha,Basar:2015ava,Aniceto:2015mto} with the lowest exponentially decaying mode present in the microscopic theory which goes beyond the stress tensor ansatz analogous to Eq.~\eqref{decompose} \footnote{For a discussion of the role played by the singularity closest to the origin in hydrodynamics obtained from kinetic theory see G.~S.~Denicol et al., Phys.\ Rev.\ D {\bf 83}, 074019 (2011).}. For the Bjorken flow in ${\cal N} = 4$ SYM this contribution was matched with the lowest non-hydrodynamic Quasinormal Mode (QNM) of AdS-Schwarzschild black brane~\cite{Heller:2013fn}. However, the presence of other singularities on the complex Borel plane remained obscured by Pad{\'e} approximation used in \cite{Heller:2013fn}, which is known to break down along branch-cuts \cite{PadeCut}. In the present case, the situation is simpler since the singularities turn out to have the structure of single poles.

%This feature allows us to see that the further singularities on the complex plane in the analytic continuation of the Borel transform of hydrodynamics come from QNMs.

%This plot is fantastic. However, due to the space limitations, we may want to consider whether just informing about the agreement is enough. It is your call.  

\begin{figure}[ht]
\includegraphics[height=0.29\textwidth]{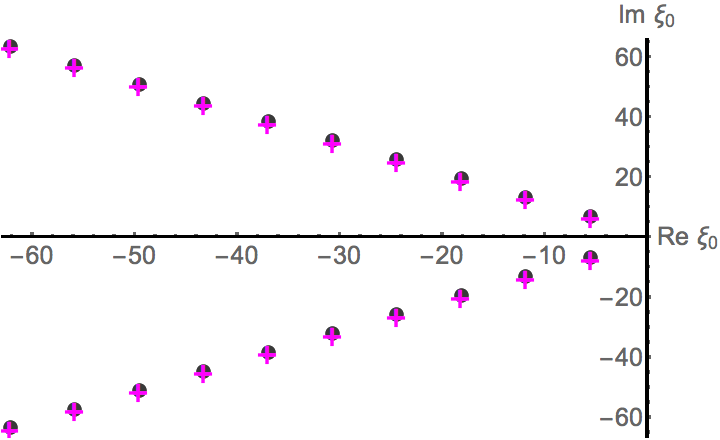}
\caption{Positions on the Borel plane of 10 singularities $\xi_{0}$ closest to the origin for 
%$\Omega_{\Delta = 2}^{(B)}$ (top) and 
$\Omega_{\Delta = 3}^{(B)}$ 
%(bottom) 
are given by solid circles. Crosses correspond to QNM frequencies for $\Delta = 3$ 
taken from Ref.~\cite{Nunez:2003eq} and redefined according to $\omega_{QNM}(T) = \hat{\omega}_{QNM} T$ and \eqref{eq.xiandomega}. 
One observes a remarkable agreement between the singularities and the QNMs.
} 
\label{fig1}
\end{figure} 

It is clear from Eq.~\eqref{eq.OmegaR} that the ambiguity in choosing the contour below or above a pole at some value of $\xi_{0}$ is the residue of the integrand. The latter (at small $g$) is
\eq
\delta \Omega_{\Delta}^{(R)} \sim e^{-\xi_{0}/g},
\label{eq.ambiguity}
\eqx
i.e. non-perturbative in $g$ at small $g$. On the other hand, the QNMs are non-hydrodynamic excitations of the microscopic theory that behave at small $g$ as~\footnote{Similarly to \cite{Heller:2013fn,Heller:2015dha,Basar:2015ava,Aniceto:2015mto} we evaluated them at zero spatial momentum.}
\eq
\delta \Omega_{\Delta}^{(QNM)} \sim e^{-i \int_{t_{0}}^{t} dt \, \omega_{QNM}(T)}
\eqx
with $\Im{(\omega_{QNM})} < 0$. It is sufficient to consider QNMs of the AdS-Schwarzschild black brane for which $\omega_{QNM}(T) = \hat{\omega}_{QNM} T$,
%\eq
%\omega_{QNM}(T) = \hat{\omega}_{QNM} T,
%\label{eq.qnmredef}
%\eqx
where $\hat{\omega}$ is a number taking discrete values. Evaluating the integral in the exponent for $ 
T= \frac{\mu}{4 \pi\,a}$ %given by Eq.~\eqref{eq.tempexpand}
gives
\eq
\delta \Omega_{\Delta}^{QNM} \sim e^{i \, \hat{\omega}_{QNM}/g}.
\label{eq.qnmeval}
\eqx
A comparison between Eqs.~\eqref{eq.ambiguity} and \eqref{eq.qnmeval} leads to
\eq
\xi_{0} = - i \, \hat{\omega}_{QNM},
\label{eq.xiandomega}
\eqx
which we checked numerically. Fig.~\ref{fig1} depicts the positions of
the first 10 singularities closest to the origin~($\xi_{0}$) of the
analytic continuation of the Borel transform of hydrodynamic series
for %$\Omega_{\Delta = 2}^{(B)}$ (i.e. for ${\cal O}_{b}$) and
$\Omega_{\Delta = 3}^{(B)}$ (i.e. for ${\cal O}_{f}$).
%$\Omega_{\Delta = 2}^{(B)}$ and $\Omega_{\Delta = 3}^{(B)}$. 
Quite
remarkably, they match at the level of a fraction of percent or better
with the frequencies of the 10 lowest non-hydrodynamic QNM frequencies
for $\Delta = 3$  operator in the AdS-Schwarzschild
background \cite{Nunez:2003eq}. Systematic improvements of the
accuracy of our calculations should allow to recover more
frequencies. Our result thus clearly demonstrates, for the first time,
that hydrodynamic theory must know about its \emph{whole} high
frequency completion via the large-order structure of the gradient
expansion. Also, the striking match between the singularities of the
analytically continued Borel transform and the QNM frequencies
obtained in \cite{Nunez:2003eq} using very different methods is a
highly nontrivial check of our calculations.

One final comment about the resurgent structure of the hydrodynamic expansion is in order. 
%Our analysis revealed QNMs of the AdS-Schwarzschild being the degrees of freedom of ${\cal N} = 4$ SYM theory. Remember, however, that we are working in the regime in which the scalar fields do not backreact the AdS-Schwarzschild geometry. 
Our calculations are performed in the regime where the scalars 
do not backreact the AdS-Schwarzschild geometry. Upon including subleading corrections to the equations of motion in the small-mass expansion we expect to recover the QNMs of ${\cal N} = 2^{*}$ plasma computed in~\cite{Buchel:2015saa}. Also, such analysis is expected to reveal additional singularities on the complex Borel plane at positive integer multiples of the ${\cal N} = 2^{*}$ QNMs, which arise from nonlinear effects in the equations of motion that we neglected.
% in our calculations.
% Fig.~\ref{fig1}

%de Sitter cosmology is obtained as a limit $\w\to -1$, in this case $\calt_{\Delta,n}[a]=\Gamma(n+4-\Delta) a^nH^n$.
%Thus, the hydrodynamic gradient expansion $\eqref{deltarate}$ has a zero radius of convergence. To go beyond the leading order hydrodynamic approximation in \eqref{deltarate}, we solve nested equations \eqref{radial} numerically.

\vspace{10 pt}

%This discussion about the non-hydro regime below is very interesting but, if one has to think about the space limitations of PRL, perhaps one can focus this paper just on the hydro limit. This discussion could appear in a long paper about this. It is your call. 

\noindent {\it Non-hydrodynamic regime: $\dot a/\mu\gg 1$}. Once more we consider de Sitter cosmology.
It is convenient in this case to introduce 
dimensionless conformal 
time $\tau=-\mu/(aH)\in (-\infty,0)$, and use the radial coordinate $z=\mu/(ar) \in (0,1)$. 
The limit $\tau\to 0$ represents a highly non-hydrodynamic (non-adiabatic) regime: $a H/\mu\gg 1$. In this limit the 
solution to  \eqref{phidelta} can be constructed by defining $v\equiv z/\tau\equiv -H/r$,  
\eq\label{fastphi}
\phi_\Delta(\tau,z)=\sum_{k=0}^\infty \tau^{4 k}G_{\Delta,k}\left(v\right)\,,\ 
v\in \left(-\frac{a H}{\mu},0\right)
\eqx
where $G_{\Delta,k}$ satisfies a complicated nested system of second order differential equations.
The $k=0$ solution can be found analytically. Restricting to $\Delta=3$ one finds
\begin{eqnarray}
G_{3,0}(v)&=& -\frac{2m_f v}{H(1+8v)^{3/2}}\biggl((1+8v)^{1/2}(1+4v)
\nonumber\\
&&+16 v^2\ln\frac{1-\sqrt{1+8v}}{1+\sqrt{1+8v}}\biggr).
\label{fast3}
\end{eqnarray}
The entropy production is determined from the $G_{3,0}$ asymptotic behavior close to the horizon; \ie  in the limit 
$v\to -\infty$: 
$G_{3,0}=-(-v)^{3/2}\sqrt{2}m_f\pi/H +\calo((-v)^{1/2})$,
which leads to 
\eq\label{nonhdroh}
\frac{d(a^3s)}{dt}=\frac{N^2}{16\pi} 
\frac{9\pi^3}{8\Gamma\left(\frac 34\right)^4}a^4 H^3 \mu\ \dd_3^2  \left(1+\calo\left(\frac{\mu}{aH}\right)\right).
\eqx  
We verified that higher $k$ corrections in \eqref{fastphi} do no spoil this result. 
The entropy production rate \eqref{nonhdroh} is completely determined by the coupling constant of the relevant 
operator ($m_f$ in this case): the information about the plasma initial state is ``lost" since the non-hydrodynamic 
regime occurs for late times of the de Sitter expansion. In this sense the late time solution  \eqref{fast3} 
realizes the universal attractor of \cite{Heller:2015dha}.  

We also note that the resummed 
entropy production rate due to the $\Delta=3$ operator enhances the corresponding leading 
hydrodynamic production rate by a factor of $(aH/\mu)\gg 1$. For general $\Delta$, we find 
\eq
\frac{d(a^3s)}{dt}\bigg|_{resummed}\ \propto \left(\frac{aH}{\mu}\right)^{2\Delta-5}\ \frac{d(a^3s)}{dt}\bigg|_{hydro}
\label{eq.resummed}
\eqx
with $aH/\mu\gg 1$. This expression deviates substantially from 
the leading hydrodynamic approximation: if the leading effect of the conformal symmetry 
breaking in plasma is due to an operator of dimension $\Delta> \frac 52$ the production rate is 
enhanced; it is suppressed otherwise.

\vspace{10 pt}

%Dissipative effects in any \emph{realistic} matter make the cosmological 
%expansion of the Universe an irreversible process resulting in entropy production. 

\noindent {\it Discussion:} In this article we studied the entropy production of the strongly coupled QGP in a FLRW Universe using a holographic model -- the strongly coupled plasma phase of ${\cal N} = 2^{*}$ gauge theory. To leading order in the hydrodynamic approximation the entropy production
in the homogeneous and isotropic expansion is governed by the bulk viscosity. Holography was used to show that subsequent corrections form a series with zero radius of convergence, see Eq.~\eqref{tncompute}. As a result, ${\cal N} = 2^{*}$ matter in FLRW provides the \emph{first}
%, second known to the date \cite{Heller:2013fn,Heller:2015dha,Basar:2015ava,Aniceto:2015mto} after the Bjorken expansion \cite{Bjorken:1982qr}, 
example of hydrodynamic flow without conformal invariance for which this property has been identified. One of the key new features on this front revealed by our investigations is the resurgent structure of the hydrodynamic gradient expansion which includes information about $\emph{all}$ the excitations of the system, see~Fig.~\ref{fig1}. 

Our results suggest that the entropy produced by the strongly coupled QGP formed in the very early Universe cannot be described by a simple hydrodynamic derivative expansion without the addition of novel non-perturbative (in the sense of the Knudsen number series) non-hydrodynamic degrees of freedom which are, however, naturally taken into account via resurgence. Also, our studies provide a partial theoretical justification of the applicability of hydrodynamics truncated at low orders of the gradient expansion in the presence of large contribution from the bulk viscous term. Such a situation was recently encountered in a holographic model of heavy-ion collisions studied in Ref.~\cite{Attems:2016tby}.

Holography is currently the only strong coupling technique that can be used to study non-equilibrium phenomena in the Universe when its temperature was close to the QCD crossover transition. It would be fascinating to use holographic techniques to investigate if such phenomena in the primordial QGP liquid could have left their imprint in the form of primordial gravitational waves.    

\vspace{10 pt}

\noindent {\bf Acknowledgments:} We would like to thank Henriette Elvang and Matthew Johnson  
for interesting discussions. 
This work was supported by NSERC through a
Discovery Grant A.B.  Research at Perimeter
Institute is supported through Industry Canada and by the Province of
Ontario through the Ministry of Research \& Innovation. JN thanks Conselho Nacional de Desenvolvimento Cient\'{\i}fico e Tecnol\'{o}gico (CNPq) and Funda\c{c}\~{a}o de Amparo \`{a} Pesquisa do Estado de S\~{a}o Paulo (FAPESP) for financial support.

\appendix

\section*{Appendix}

\noindent {\it Microscopics of the model.---} ${\cal N} = 2^{*}$ gauge theory is obtained as a deformation of the maximally supersymmetric $SU(N)$ Yang-Mills theory. The field content of $\caln=4$ SYM theory includes the
gauge field $A_\mu$, four Majorana fermions $\psi_a$, and three complex
scalars $\phi_i$, all in the adjoint
representation. SYM theory can be deformed by adding two
independent `mass' terms \cite{Hoyos:2011uh}
 \eq\label{lagdef}
\delta \call= -2\,\int d^4x\,\left[ \,m_b^2\,\calo_b
+m_f\,\calo_f\,\right]
 \eqx
where
\begin{eqnarray}
\calo_b&=&\frac13 {\tr}\left(\, |\phi_1|^2 + |\phi_2|^2 - 2\,|\phi_3|^2
\,\right)\,,
 \nonumber\\
\calo_f&=& -{\tr}\biggl( i\,\psi_1\psi_2 -\sqrt{2}g_\mt{YM}\,\phi_3
[\phi_1,\phi_1^\dagger] +\sqrt{2}g_\mt{YM}\,\phi_3
[\phi_2^\dagger,\phi_2]  \nonumber\\
&&+ {\rm h.c.}\biggr)
 +\frac23 m_f\, {\tr}\left(\, |\phi_1|^2 + |\phi_2|^2 +
|\phi_3|^2\, \right)\,.
 \label{obof}
\end{eqnarray}
The relevant 
deformation \eqref{lagdef} breaks scale invariance and, when $m_b=m_f$, half of the supersymmetries of the parent SYM;
for general mass parameters $m_b\ne m_f$ supersymmetry is completely broken.   
In the planar limit and for large 't Hooft coupling, $\caln=2^*$ gauge theory possesses a holographically 
dual gravitational description \cite{Pilch:2000ue,Buchel:2000cn} which provides an opportunity 
to study its thermodynamic \cite{Buchel:2003ah,Buchel:2007vy}, 
hydrodynamic \cite{Buchel:2004hw,Benincasa:2005iv,Buchel:2007mf,Buchel:2008uu}, and far from equilibrium properties
\cite{Buchel:2012gw,Buchel:2015saa}. The duality between $\caln=2^*$ gauge theory at strong coupling 
and the gravitational Pilch-Warner effective action \cite{Pilch:2000ue} (PW) allows for precision tests of the holographic correspondence in a nonconformal setting \cite{Buchel:2013id,Bobev:2013cja}. 

\vspace{10 pt}

\noindent {\it Holographic equations of motion.---} Under the assumptions of homogeneity and isotropy, we obtain the following 
equations of motion, describing dynamics of $\caln=2^*$ gauge theory at strong coupling within the dual gravitational framework:
\begin{eqnarray}\nonumber
&&0=d_+'\Sigma+2 {\Sigma'}\ d_+\ln\Sigma+
\frac \Sigma6\ V,\\ \nonumber
&&0=A''-6(\ln\Sigma)'\ d_+\ln\Sigma +4\chi' d_+\chi +12\a' d_+\a 
-\frac V6 \\ \nonumber
&&d_+'\a+\frac{3}{2}\ \left((\ln\Sigma)'d_+\a+\a' d_+\ln\Sigma \right) 
-\frac {1}{48} \del_\a V, \\ \label{ev}
&&0=d_+'\chi+\frac{3}{2}\ 
\left((\ln\Sigma)'d_+\chi+\chi' d_+\ln\Sigma \right)-\frac{1}{16}\del_\chi V,
\label{ev1}
\end{eqnarray}
as well as the Hamiltonian constraint equation:
\eq\label{ham}
0=\Sigma''+\left(4 (\a')^2+\frac 43 (\chi')^2\right) \Sigma
\eqx
and the momentum constraint equation:
\begin{eqnarray}\nonumber
0&=&d^2_+\Sigma -2 A\Sigma' -(4 A \Sigma'+A' \Sigma)d_+\ln\Sigma
\\
&&+\left(4 (d_+\a)^2+\frac 43 (d_+\chi)^2\right)\Sigma  
-\frac 13 \Sigma A V.
\label{mom}
\end{eqnarray}
In Eqs. \eqref{ev}-\eqref{mom} 
we denoted $'= \frac{\del}{\del r}$, $\dot\ =\frac{\del}{\del t}$. 
%and $d_+= \frac{\del}{\del t}+A \frac{\del }{\del r}$. 
The initial state of the gauge theory is specified by providing the scalar profiles 
$\a(0,r)$ and $\chi(0,r)$ and solving the constraint \eqref{ham} 
subject to the boundary conditions given by Eq.~(9) from the article. Eqs.~\eqref{ev} can then be used 
to dynamically evolve the state.

\vspace{10 pt}

\noindent {\it $\caln=2^*$  in the conformal limit.---}
When $m_b=m_f=0$, the case of the parent conformal SYM, 
equations of motion 
\eqref{ev1}-\eqref{mom}
yield the \emph{exact} solution
\eq
\a=\chi=0\,,\ \Sigma=\frac {ar}{2}\,,\ A=\frac{r^2}{8}\left(1-\frac{\mu^4}{r^4a^4}\right)-\frac {\dot a}{a}\ r,
\label{sym}
\eqx
where the \emph{dimensionful} constant $\mu$ is related to the local temperature~via
\eq
%\mu=4\pi aT.
T= \frac{\mu}{4 \pi\,a}.
\label{eq.tempexpand}
\eqx
We find in this case, in agreement with \cite{Apostolopoulos:2008ru},
\eq\label{cftres}
\e=\frac 38\pi^2 N^2 T^4+\frac{3N^2(\dot a)^4}{32\pi^2a^4}\,,\ P=\frac 13\e
-\frac{N^2(\dot a)^2\ddot a}{8\pi^2 a^3}.
\eqx
It is clear (see \cite{Apostolopoulos:2008ru}) that the stress tensor 
\eqref{cftres} arises from a conformal transformation performed on an equilibrium state in Minkowski while redefining the time variable to bring the background metric to take the FLRW form.

\bibliography{references}

\end{document}